\documentclass[reprint,twocolumn,secnumarabic,amssymb, nobibnotes, aps, pre,showkeys]{revtex4-2}
\usepackage{bm,chemformula,graphicx,titlesec}
\usepackage[unicode]{hyperref}
\hypersetup{colorlinks= true,
	citecolor= blue,
	linkcolor= red,
	urlcolor=magenta,
	filecolor=cyan,
	linkbordercolor={1 0 0},
	citebordercolor={0 1 0},
	urlbordercolor={0 1 1},
} 
\usepackage[mathlines]{lineno}
\usepackage{tikz}
\usepackage{float,wrapfig}
\usepackage{microtype}
\usepackage[tight]{subfigure}
\DeclareUnicodeCharacter{2212}{-}
\begin{document}
	\preprint{APS/123-QED}
	\title{Nonequilibrium thermodynamic characterization of chimeras in a continuum chemical oscillator system}
	\author{Premashis Kumar}
	\affiliation{S. N. Bose National Centre For Basic Sciences, Block-JD, Sector-III, Salt Lake, Kolkata 700 106, India}
	\author{Gautam Gangopadhyay}
	\email{gautam@bose.res.in}
	\affiliation{S. N. Bose National Centre For Basic Sciences, Block-JD, Sector-III, Salt Lake, Kolkata 700 106, India}
	\date{\today}
	
	\begin{abstract}
		The emergence of the chimera state as counterintuitive spatial coexistence of synchronous and asynchronous regimes is addressed here in a continuum chemical oscillator system by implementing a relevant complex Ginzburg-Landau equation with global coupling. This study systematically acquires and characterizes the evolution of nonequilibrium thermodynamic entities corresponding to the chimera state. The temporal evolution of the entropy production rate exhibits a beat pattern with a series of equidistant spectral lines in the frequency domain. Symmetric profiles associated with the incoherence regime appear in descriptions of the dynamics and thermodynamics of the chimera. It is shown that identifying the semigrand Gibbs free energy of the state as the Gabor elementary function can unveil the guiding role of the information uncertainty principle in shaping the chimera energetics.
	\end{abstract}
	\maketitle
	
	\section{\label{sec1}Introduction}
	Collective dynamics of systems can exhibit emergent behaviors with remarkably rich and complex features across all scales. In diverse contexts, extensive investigations have been carried out to understand these behaviors and recognize their applicability. One of such emergent phenomena that has recently received widespread scientific interest from different fields due to its intriguing and subtle nature is the chimera state~\citep{kuramoto2002coexistence, strogatz}. chimera~\citep{kuramoto2002coexistence, strogatz}, a counterintuitive state in the collective dynamics of coupled identical entities or continuous media~\citep{mcglelocalglobal, chimeracontinuousmedia}, emerges as a  coexistence of spatially coherent and incoherent behavior. chimera state has received widespread scientific interest due to its intriguing and subtle nature and presence in various theoretical and experimental frameworks~\citep{PARASTESH20211, Sindre2021}. Initially, it was believed that chimera could only exist in the array of identical phase oscillators connected by a nonlocal coupling. However, later on, evidence of ``amplitude-mediated chimeras"~\citep{SethiaSen1} in the nonlocal coupling version of the complex Ginzburg-Landau equation(CGLE)~\citep{Kuramoto1984ChemicalTurbulence, aranson2002world, Cross2009PatternSystems} establishes the connection of this peculiar state to the amplitude dynamics of the system. Furthermore, chimera state was also realized in the global(all-to-all coupling) coupling~\citep{schmidtglobal, sethiasen} scheme and in purely locally coupled networks~\citep{Lainglocal}. All these key advancements broaden the scope of chimera investigation into more diverse settings of collective dynamics with the main focus dedicated to a better understanding of chimera dynamics in the different coupling schemes with varied coupling strenth and various types of oscillator collections.  
	
	Nevertheless, one of the first experimental evidence of chimera was reported in a population of nonlocally coupled discrete photosensitive Belousov-Zhabotinsky(BZ) chemical oscillators~\citep{Tinsleynature2012}. In nonlocally coupled chemical oscillators~\citep{nkomo2016}, different variants of chimera have been more recently demonstrated and characterized both experimentally and theoretically. However, previous works on chemical systems with global coupling have only observed oscillatory cluster patterns~\citep{Epsteinoscillatorycluster, bzvanag} or turbulent state~\citep{cgleglobal}. This study aims to generate the chimera state in a simple prototypical chemical oscillator system using the global coupling scheme. The emergence of chimera within the globally coupled framework of chemical oscillator would complement the investigation of various possible patterns in a chemical system. 
	
	More importantly, all previous studies regarding chimera states have been limited to the dynamic aspects of the state. However, the occurrence of a similar state in neuroscience~\citep{chowbumps, RATTENBORG2000817}, hydrodynamics~\citep{barkley, Duguet} and possible association of the chimera with different brain states, for example, unihemispheric sleep~\citep{Abrams2015} in different aquatic and avian species, seek a complete thermodynamic description of such state to shed light on basic underlying mechanisms and signatures of these similar states. A proper thermodynamic description of such emergent behaviors is still lacking. We have focused on this aspect by characterizing the chimera state in terms of thermodynamic entities. This thermodynamic investigation of a peculiar state like the chimera can broaden the current understanding of the coexistence of qualitatively different regimes and transitions among them. In this respect, the chemical work needed to manipulate such a state and the efficiency of information spreading in the presence of the state are knowledge of crucial importance. Besides theoretical understandings, the thermodynamic picture can serve useful purposes in potential applications of such situations or exploring the real-world relevance of these states.
	
	This study will realize the chimera state in a prototypical continuum chemical oscillator system by imposing a nonlinear globally coupled version of CGLE and then systematically characterize the state by implementing a suitable nonequilibrium thermodynamic framework. Instead of direct incorporation of the global coupling in the reaction-diffusion system(RDS) of the Brusselator, we have opted to generate the effect of coupled dynamics within the system using the modified CGLE(MCGLE)~\citep{mcgle1st, mcgle2, schmidtglobal, mcglelocalglobal, Haugland2015SelforganizedAC} with the global coupling at the linear and nonlinear levels. The chimera state concentration yielding the nonequilibrium nature of the system will be obtained by inserting numerically obtained amplitude from the MCGLE into an analytical equation of the concentration dynamics. Our approach thus facilitates analytically tractable thermodynamics within the nonequilibrium framework utilized here. 
	
	The layout of the paper is as follows. First, we have described the dynamics of the Brusselator reaction-diffusion system in sec.~\ref{secII}. In the next section, the Brusselator system in the presence of global coupling is represented in terms of a relevant amplitude equation, and concentration fields of the intermediate species are obtained. In sec.~\ref{secIV}, we have formulated the entropy production rate and the semigrand Gibbs free energy to present the nonequilibrium thermodynamic picture. Then, we have provided results and discussions in sec.~\ref{secV}. Finally, the paper is concluded in sec.~\ref{secVI}.
	
	\section{\label{secII}Dynamics of Brusselator RDS}
	The Brusselator model~\citep{Prigogine1968SymmetryII, Nicolis1977Self-organizationFluctuations}, a minimal abstract model in chemical kinetics, is capable of capturing self-sustained oscillatory behavior in Belousov-Zhabotinsky reaction~\citep{Zhabotinsky1991AWaves} and various other chemical and biological systems and has been exploited extensively for investigating many intricate and cooperative behaviors. The reversible Brusselator model is described through the following chemical reactions: 
	\begin{equation}
		\begin{aligned}
			\rho&=1:&\ch{A&<=>[\text{k\textsubscript{1}}][\text{k\textsubscript{-1}}] X}\\
			\rho&=2: &\ch{B + X&<=>[\text{k\textsubscript{2}}][\text{k\textsubscript{-2}}]Y + D}\\
			\rho&=3:& \ch{2 X + Y&<=>[\text{k\textsubscript{3}}][\text{k\textsubscript{-3}}]3X} &\textsf{(Autocatalytic)}\\ 
			\rho&=4:&\ch{X&<=>[\text{k\textsubscript{4}}][\text{k\textsubscript{-4}}]E}
		\end{aligned}
		\label{crn}
	\end{equation}
	with $'\rho'$ being reaction step label. This reaction network consists of two types of species: intermediate species with dynamic concentration, $\{X,Y\}\in I$ and externally controllable chemostatted species with a constant homogeneous concentration within the time scale of interest, $\{A, B, D, E\}\in C$. Thus, the concentration dynamics of intermediate species of the Brusselator obeys following rate equations under the assumption that forward  reaction  rate  constants $k_{\rho}$ are much higher than reverse ones,
	\begin{equation}
		\begin{aligned}
			\dot{x}&={k_1}a-({k_2}b+k_4)x+{k_3}x^2y \\
			\dot{y}&={k_2}bx-{k_3}x^2y 
			\label{dynamic}
		\end{aligned}
	\end{equation}
	with $x=[X],y=[Y],b=[B],a=[A]$ denoting concentrations of species. Consequently, steady-state values of intermediate species concentration are acquired as,
	$x_{0}=\frac{k_1}{k_4}a, y_{0}=\frac{{k_2}{k_4}}{{k_1}{k_3}}\frac{b}{a}$ and the Jacobian matrix, $\mathcal{J}$ is extracted from eq.~\eqref{dynamic} with elements $J_{11}=-({k_2}b+k_4)+2{k_3}x_0y_0$, $J_{12}={k_3}{x_0}^2$, $J_{21}={k_2}b-2{k_3}x_0y_0$ and  $J_{22}=-{k_3}{x_0}^2$. Now as the control parameter, $b$ is gradually varied, the Hopf instability would arise under the requirement, $J_{11}+J_{22}=0$ at the onset, and hence the critical value of the control parameter is, $b_{cH}=\frac{k_{4}}{k_{2}}+\frac{k_{1}^2 k_{3}}{k_{2}{k_4}^2}a^2$. The critical eigenvector, $U_{cH}$  corresponding to the largest eigenvalue, $\lambda_{+}=i \sqrt{\frac{k_{1}^2 k_{3}}{k_4}}a$, is
	$U_{cH}=
	\begin{pmatrix}
		1+\frac{i}{a}\sqrt{\frac{k_4}{k_3}}\frac{1}{k_1}&,
		-(1+\frac{{k_4}^3}{k_3 {k_1}^2}\frac{1}{a^2}) 
	\end{pmatrix}^\textbf{T}$. Additionally, the critical frequency of the oscillation for Hopf instability, $f_{cH}$ can be obtained from the imaginary part of the eigenvalue at the onset of instability. Therefore, the oscillation frequency near the the Hopf instability is approximately, $f_{cH}=\sqrt{\frac{k_{1}^2 k_{3}}{k_4}}a$ for the Brusselator model. 
	
	After taking diffusion into account, the Brusselator RDS in one spatial dimension $r\in [0,l]$ reads
	\begin{equation}
		\begin{aligned}
			\dot{x}&={k_1}a-({k_2}b+k_4)x+{k_3}x^2y+D_{11}x_{rr}\\
			\dot{y}&={k_2}bx-{k_3}x^2y+D_{22}y_{rr}, 
			\label{ddynamic}
		\end{aligned}
	\end{equation}
	with $D_{11}$, $D_{22}$ being constant self-diffusion coefficients of intermediate species $X$ and $Y$, respectively. The Jacobian matrix of the RDS in eq.~\eqref{ddynamic} is $\mathcal{J_D}=\mathcal{J}-q^2\mathcal{D}$ with $q$ being the wavenumber. In the presence of diffusion, the onset of Hopf instability demands Tr$(\mathcal{J_{D}})=0$, and hence the critical value of the control parameter, $b$ is specified as,
	$b_{ctw}=\frac{k_{4}}{k_{2}}+\frac{k_{1}^2 k_{3}}{k_{2}{k_4}^2}a^2+\frac{(D_{11}+D_{22})}{k_2}q^2$,
	where the wavenumber, $q=\frac{2n\pi}{l}$ according to periodic boundary conditions in the finite domain, $l$ with $n$ being an integer. However, to restrict our investigation solely to the Hopf instability regime, we have set the wavenumber, $q=0$, here. 
	\section{\label{secIII}Brusselator Representation with Global Coupling}
	The amplitude dynamics encapsulates the essential role of nonlinearity in pattern formation~\citep{Cross2009PatternSystems}. For a nonlinear chemical system like Brusselator, the concentration dynamics of the intermediate species can be acquired by exploiting the amplitude as the multiplicative factor in the standard linear stability description of the system. Here we have considered that a globally coupled system of Brusselators can be effectively represented in terms of the amplitude of MCGLE near the onset of Hopf instability since the amplitude of CGLE guides reaction-diffusion dynamics near the Hopf bifurcation point by capturing crucial nonlinear features of the system. The normal form of the CGLE~\citep{Nicolis1995IntroductionScience, Cross2009PatternSystems} in spatially extended system can be expressed as  
	\begin{equation}
		\frac{\partial Z}{\partial t}=\lambda Z -(1-i\beta)\mid Z \mid ^2Z+(1+i \alpha)\partial_{r}^2 Z.
		\label{ncgle}
	\end{equation}
	with $Z$ being the amplitude field and $\lambda$, $\beta$ and $\alpha$ being coefficients encompassing the details of a particular system.  In the presence of a global coupling, the normal form of the CGLE in eq. \eqref{ncgle} can be recast into the MCGLE~\citep{mcgle1st,mcgle2}, 
	\begin{eqnarray}
		\frac{\partial Z}{\partial t}=\lambda Z -(1-i\beta)\mid Z \mid ^2Z+(1+i \alpha)\partial_{r}^2 Z \nonumber \\                     
		-(\lambda+i\nu) \left\langle Z \right\rangle
		+(1-i\beta)\left\langle \mid Z\mid ^2Z \right\rangle
		\label{gcncgle}
	\end{eqnarray}
	where $\left\langle...\right\rangle$ denotes the spatial average. Now spatial average over the eq.~\eqref{gcncgle} yields an oscillatory mean field, $\left\langle Z \right\rangle=Z_0=\eta \exp(-i\nu t)$ with $\eta$ and $\nu$ being the amplitude and the frequency of the oscillation, respectively. This mean-field oscillation is a feature of the nonlinear global coupling of the MCGLE. By substituting $Z=Z_0(1+Z_{IH})$, with an arbitrary inhomogeneity $Z_{IH}$, into eq.~\eqref{gcncgle} followed by a linear stability analysis, one can determine the threshold value of the $\eta$ as $\eta_c=\sqrt{\frac{\lambda}{2}}$, below which uniform oscillation becomes unstable irrespective of other parameter values. Further, coefficients $\alpha$ and $\beta$ are obtained by implementing Krylov-Bogolyubov(KB) averaging method~\citep{krylov1949introduction, pkgg} to normal form of CGLE. For the Brusselator, coefficients are acquired as
	$\alpha=\frac{\Omega(D_{22}-D_{11})}{(D_{11}+D_{22})}$, 
	$\beta=\frac{p_2}{p_1}\frac{1}{3a}$, and  $\lambda =\frac{b-1-a^2}{2}$ with the ratio of correction factors, $\frac{p_1}{p_2}$ being $\frac{4-7a^2+4a^4}{(2+a^2)}$. For the investigation of the chimera state, we have set $\eta=0.66\sqrt{\lambda}$ and $\nu=\frac{\lambda}{10}$ throughout this report.
	
    To acquire the amplitude in the presence of the global coupling, we have solved eq.~\eqref{gcncgle} numerically using a pseudospectral method incorporated with an exponential time differencing algorithm~\citep{COX2002}. For the simulation purpose, we have exploited a computational timestep of 0.01 and have considered 2048 grid points for the spatial domain used in this investigation. As an initial state of the system, a uniform state with additional noise has been chosen. We have applied periodic boundary conditions. In the numerical simulation, all coefficients of eq.~\eqref{gcncgle} have been specified in terms of parameters of the Brusselator in eq.~\eqref{ddynamic}. On combining the numerically obtained amplitude field with linear stability description of the nonlinear system, the  collective concentration dynamics of the Brusselator system  has been acquired through the following equation,
	\begin{equation}
		{z_I}_{H}={z_I}_{0}+A_{M}U_{cH}\exp(i f_{cH}t)+C.C.
		\label{hwave}
	\end{equation} 
	with $A_{M}$ being the numerically derived amplitude field from the eq.~\eqref{gcncgle}, ${z_I}_{0}$ being the initial uniform concentration field set by steady-state values of two intermediate species, and $f_{cH}=\sqrt{\frac{k_{1}^2 k_{3}}{k_4}}a$ being the critical frequency of the Brusselator within the Hopf instability regime. Here, the chimera state arises in the chemical system in the presence of global coupling, and we have considered this coupling to be present at the level of amplitude dynamics only. Thus incorporation of numerically obtained amplitude with the linear stability representation of the Brusselator system would illustrate the essential feature of the chimera in the concentration dynamics.
	
	\section{\label{secIV}Nonequilibrium Thermodynamic Description}
	We have employed a recently developed nonequilibrium thermodynamic framework~\citep{Rao2016NonequilibriumThermodynamics, Falasco2018InformationPatterns} to capture the entropic and energetic description of the chimera state. Incorporation of the nonequilibrium steady state representation with the MCGLE scheme makes the thermodynamic description of the collective system possible.
	\subsection{Entropy Production Rate}
	The entropy production rate(EPR) due to the chemical reaction can be derived by utilizing flux-force form in which fluxes obey the mass action law, $j_{\pm \rho}=k_{\pm\rho}\prod_{\sigma}z^{v_{\pm\rho}^{\sigma}}_{\sigma}$ with $'+'$ and $'-'$ label denoting the forward and backward reaction, respectively and the force is affinity of reaction~\citep{Prigogine1954ChemicalDefay.}, $f_{\rho}=-\sum_{\sigma}{S_{\rho}^{\sigma}\mu_{\sigma}}$ with $S_{\rho}^{\sigma}=v_{{-}\rho}^{\sigma}-v_{{+}\rho}^{\sigma}$ being the stoichiometric coefficient of species and  $\mu_{\sigma}$ being the chemical potential. For the solvent concentration $z_0$  and the standard-state chemical potential,$\mu_{\sigma}^o$, the chemical potential is given as $\mu_{\sigma}=\mu_{\sigma}^o+\ln{\frac{z_{\sigma}}{z_0}}$. Thus, the entropy production rate due to the chemical reaction is given as
	$\frac{d\Sigma_{R}}{dt}=\frac{1}{T}\int  dr \sum_{\rho} (j_{+\rho}-j_{-\rho}) \ln{\frac{j_{+\rho}}{j_{-\rho}}}$ with affinities being expressed in terms of the reaction fluxes, $f_{\rho}= \ln{\frac{j_{+\rho}}{j_{-\rho}}}$, and $T$ being the constant absolute temperature set by solvent. Similarly, diffusive flux and affinity represent the entropy production rate due to diffusion as, 
	$\frac{d\Sigma_{D}}{dt}=\int dr \Big[ D_{11}\frac{{\parallel{\frac{\partial x}{\partial r} }\parallel}^2}{x}+D_{22}\frac{{\parallel{\frac{\partial y}{\partial r} }\parallel}^2}{y}\Big] \label{eprdd}$,
	with the total entropy production rate having comprised of reaction entropy production rate and diffusion entropy production rate.
	
	\subsection{Semigrand Gibbs Free Energy}
	Now the stoichiometric matrix of the chemical reaction network in eq. \eqref{crn} is
	\begin{gather}
		S_{\rho}^{\sigma}=
		\bordermatrix{ ~ & R_{1} & R_{2}&R_{3}&R_{4}\cr
			X&1 &-1&1&-1\cr
			Y&0&1&-1&0 \cr
			A&-1&0&0&0 \cr
			B&0&-1&0&0\cr
			D&0&1&0&0 \cr
			E&0&0&0&1\cr} .\label{st}
	\end{gather}
	In a closed system, conservation laws~\citep{Alberty2003ThermodynamicsReactions} are specified as the left null vectors corresponding to left null space of the stoichiometric matrix, $\sum_{\sigma}{l_{\sigma}^{\lambda}S_{\rho}^{\sigma}}=0$, where $\{l_{\sigma}^{\lambda}\}\in \mathbb{R}^{(\sigma-w )\times \sigma}, w=rank(S_{\rho}^{\sigma}).$ Components, globally conserved quantities of the reaction network can be constituted using conservation laws as $L_{\lambda}=\sum_{\sigma}{l_{\sigma}^{\lambda}}z_{\sigma}$ such that $\frac{d}{dt}\int dr L_{\lambda}=0$. For the stoichiometric matrix in eq. \eqref{st}, the conservation laws of the closed  reaction network are acquired as following  two linearly independent $(1\times 6)$ vectors, $l_{\sigma}^{\lambda=1}=  
	\bordermatrix{~&X&Y&A&B&D&E\cr
		&1&1&1&0&0&1\cr}$  and   $l _{\sigma}^{\lambda=2}=   
	\bordermatrix{~&X&Y&A&B&D&E\cr
		&0&0&0&1&1&0\cr}.$
	Hence, the corresponding components are expressed as, $L_1=x+y+a+e$ and $L_2=b+d$. A conservation law can be broken when a closed system is opened by chemostating. Therefore, conservation laws in an open system are generally characterized as $\{l^{\lambda}\}=\{l^{\lambda_b}\}\cup\{l^{\lambda_u}\}$, and the set of chemostatted species can be divided into two subsets $\{C\}=\{C_{b}\}\cup\{C_{u}\}$  with labels $u$ and $b$ denoting unbroken and broken ones, respectively. Here, both conservation laws of the Brusselator model can be broken by chemostatted species $A$ and $B$. Therefore, the pair, $A$ and $B$ belong to the set $C_{b}$ and have been considered as the reference chemostatted species here.
	
	In terms of the chemical potential, the nonequilibrium Gibbs free energy of a reaction network is defined as~\citep{Fermi1956Thermodynamics} $G=G_{0}+ \sum_{\sigma \neq 0}{(z_{\sigma}\mu_{\sigma}-z_{\sigma})}$ with $G_{0}=z_{0}\mu_{0}^o$. However, the Gibbs free energy do not represent the proper energetics of the system operating far from equilibrium. To capture the  proper energetics of the chimera, we need to employ the semigrand Gibbs free energy(SGG) obtained by the Legendre transformation of the nonequilibrium Gibbs free energy~\citep{Rao2016NonequilibriumThermodynamics, Falasco2018InformationPatterns}, 
	\begin{equation}
		\mathcal{G}=G-\sum_{\lambda_b}{\mu_{\lambda_b}M_{\lambda_b}}
	\end{equation} 
	with $M_{\lambda_b}=\sum_{C_b}l_{C_b}^{{\lambda_b}^{-1}} L_{\lambda_b}$ being moieties~\citep{Haraldsdttir2016IdentificationOC} exchanged between chemostats and system. 
	
	\section{\label{secV}Results and Discussions}
	\begin{figure*}[t]
		\includegraphics[width=\textwidth]{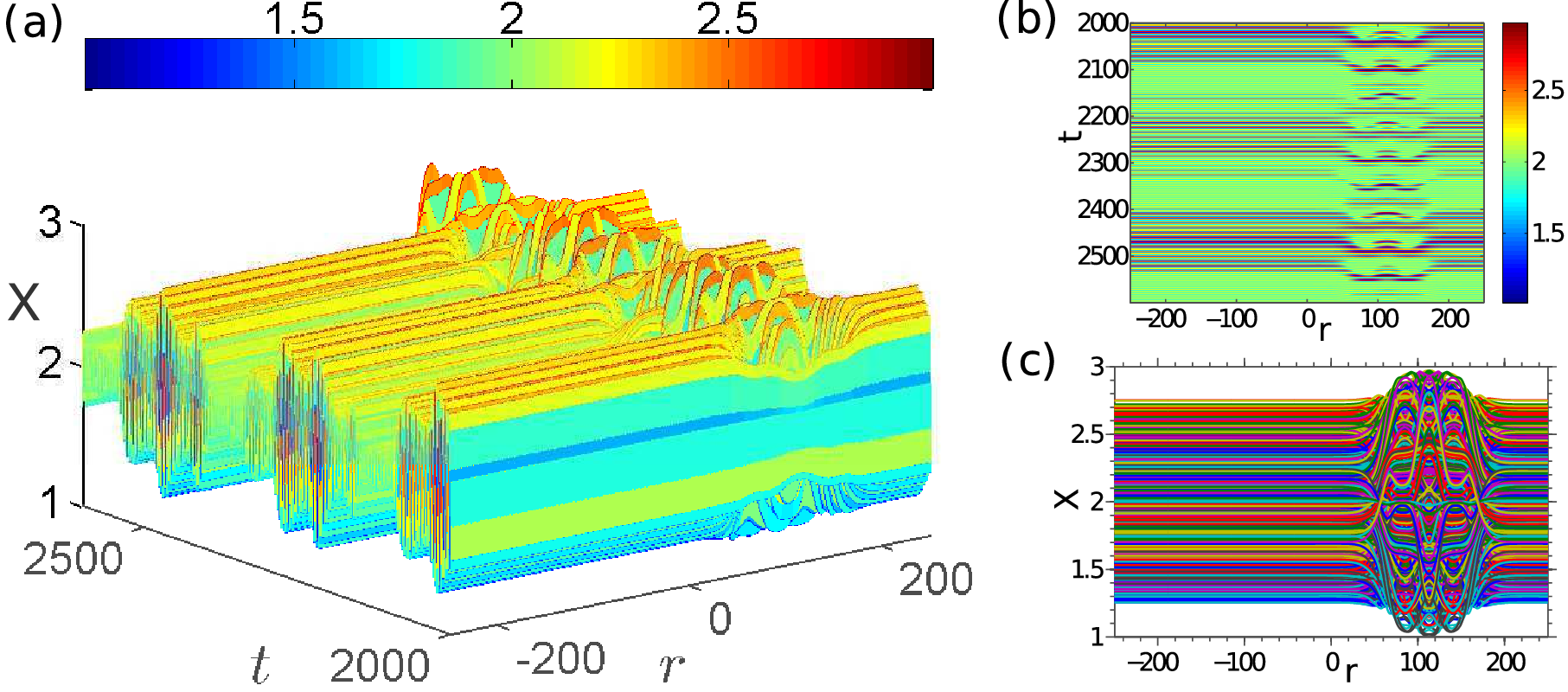}
		\caption{\label{concentration} (a) The 3D concentration field, and (b) corresponding space-time image of the activator in the Brusselator system. (c) The concentration dynamics along the spatial axis. The concentration dynamics has been explicitly assessed from eq.~\eqref{hwave} by exploiting the $2001$ amplitude snapshots between time $t=2000$ and $t=2600$. For the simulation of the related MCGLE, we have used a timestep size of $0.01$ and divided the one-dimensional system length, $l=500$, into 2048 grid points, and started from a uniform state at $t=0$. All illustrations are obtained for $D_{11}=4, D_{22}=3.2, l=500, a=2, b=5.28,\text{and }   k_{-{\rho}} =10^{-4}\ll k_{\rho} = 1$.}
	\end{figure*}
	
	The spatiotemporal evolution of the activator concentration of the system is depicted in fig.~\ref{concentration}. Initially, the system is kept at a uniform base state. The chimera concentration dynamics has been obtained from eq.~\eqref{hwave} by using the $2001$ amplitude snapshots between time $t=2000$ and $t=2600$. For the simulation of the MCGLE, we have considered a timestep size of $0.01$ and divided the one-dimensional system of length  $l=500$ into 2048 grid points. All illustrations are obtained for diffusion coefficients $D_{11}=4, D_{22}=3.2$, chemostatted concentrations $a=2, b=5.28$, and chemical reaction rate constants, $k_{-{\rho}} =10^{-4}\ll k_{\rho} = 1$. The appearance of the chimera state in the 3D concentration field of the activator is observed in fig.~\ref{concentration}(a). The corresponding space-time realization of the chimera is presented in fig.~\ref{concentration}(b). The image shows an embedded incoherence regime within the coherence counterparts of the state. The incoherence nature of the activator concentration field is apparent in fig.~\ref{concentration}(c) with the turbulent concentration around $r\approx 100$. It is also realized from the concentration field that the coherence region evolves with time.
	\begin{figure*} 
		\includegraphics[width=\textwidth]{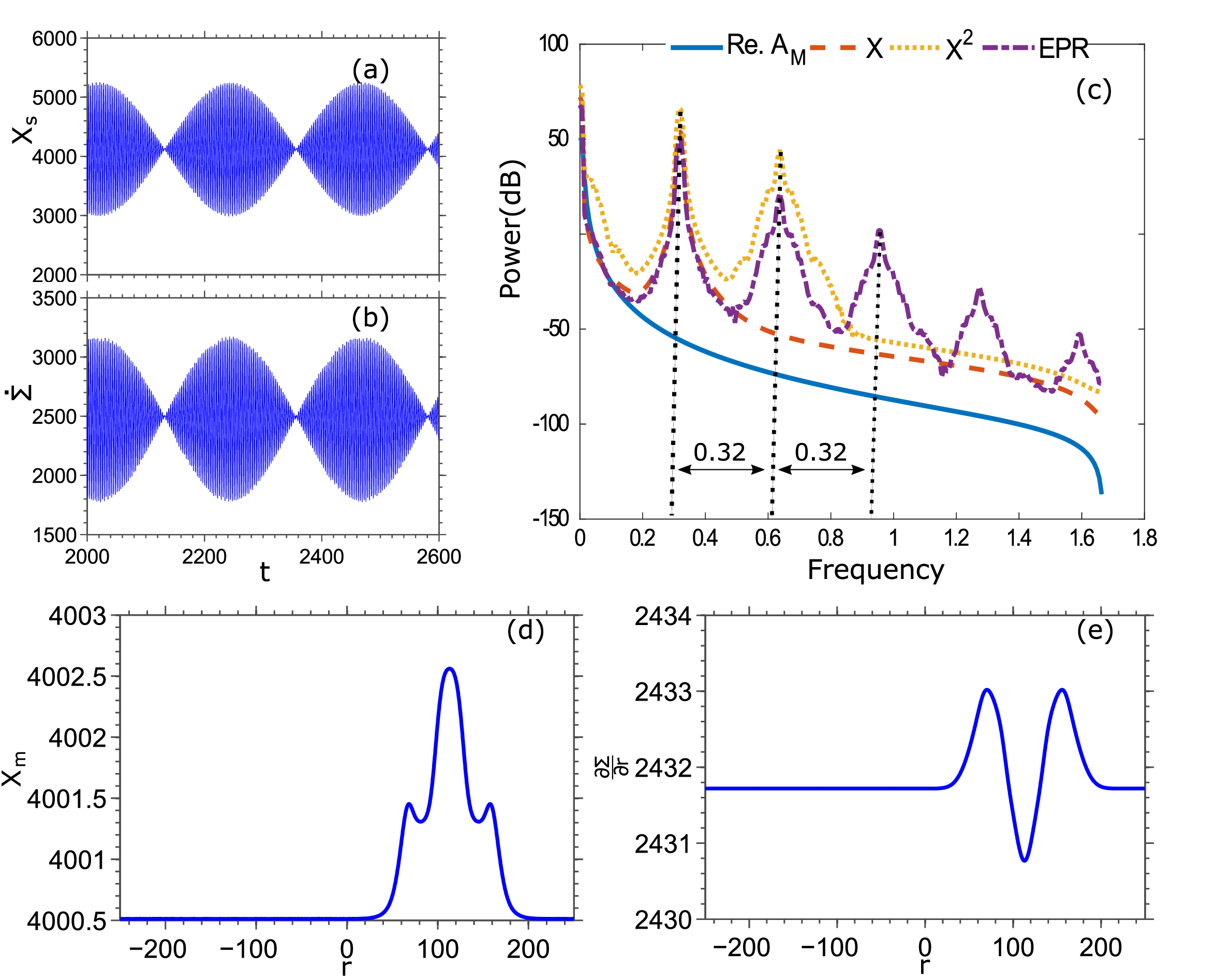}
		\caption{\label{globalepr} (a) Space-integrated concentration, $X_s$ variation with time shows a beat interference pattern. (b) The Entropy production rate(EPR) has similar qualitative behavior as temporal dynamics of global concentration. (c) The power spectrum of the space-integrated amplitude, concentration, and EPR. For power spectrum, we have implemented Welch's power spectral density estimate using a Hann window. (d) Time-integrated concentration, $X_m$ over the spatial axis. (e) The spatial EPR. In (d) and (e), the incoherence state can be identified as symmetric profiles. Here, space-integrated concentration, $X_s$, and time-integrated concentration, $X_m$ are obtained by taking the sum over the spatiotemporal concentration field of intermediate species $X$ along the spatial length and temporal domain, respectively. For Chemostatted species concentrations are set at $b=5.28, a=2,$ and $d=e=10^{−4}$ with $d, e$ denoting the concentration of $D, E$, respectively.}
	\end{figure*}
	We have acquired space-integrated concentration, $X_s$, and time-integrated concentration, $X_m$ from the spatiotemporal concentration field of intermediate species $X$ in  eq.~\eqref{hwave} by taking the sum over the spatiotemporal concentration data along the spatial length and temporal domain, respectively. The temporal evolution of space-integrated concentration, $X_s$ in fig.~\ref{globalepr}(a) reveals a beat interference pattern with a periodic envelope of the temporal profile. A similar beat phenomenon in the chemical system of periodically forced pH oscillator~\citep{beatphoscillator} has been recently reported. We assert that the emergence of a beat pattern in the temporal domain of the chimera state is associated with the global coupling scheme of the amplitude equation. The temporal trait of the entropy production rate in fig.~\ref{globalepr}(b) seemingly reflects qualitatively the same behavior as the space-integrated concentration in fig.~\ref{globalepr}(a). However, in fig.~\ref{globalepr}(c), the power spectrum of the space-integrated concentration has two lines, whereas the EPR power spectrum exhibits a series of equidistant spectral lines with gradually decreasing power in the frequency domain. These frequencies related to the spectral lines are determined by a center frequency, $f_o$ and spacing between consecutive lines, $\delta f$  as $f_n = n(\delta f) + f_{o}$ with n being an integer. This feature in the frequency domain of the entropy production rate is generated due to the underlying mixing of frequency components of concentrations in the entropy production rate representation and the intrinsic nonlinearity of the system. The generation of new frequencies is evident from the emergence of an additional frequency line in the power spectrum of the space-integrated $X^2$.Power spectra in fig.~\ref{globalepr}(c) are obtained by implementing Welch's power spectral density estimate~\citep{Welch} using a Hann window and an adequate discrete Fourier transform(DFT) length in MATLAB.
	
	Now, the time-integrated activator concentration in fig.~\ref{globalepr}(d) and the spatial EPR in fig.~\ref{globalepr}(e) have an even-symmetric structure corresponding to the incoherence state of the chimera in an otherwise flat profile associated with the coherence regimes. This symmetry of the incoherence state in dynamic and thermodynamic entities possibly inherits from the uniform nature of the global coupling. The time-integrated concentration has a composite pulse structure in the incoherence regime with an increased concentration at the axis of symmetry. Away from the axis, the principal concentration peak of the pulse subsides into a secondary peak preceded by a notch on both sides. The incoherence state in the spatiotemporal realization of the chimera can be visualized as the inclusion of strong spatial amplitude fluctuation over time to this symmetric time-integrated profile. The spatial EPR in fig.~\ref{globalepr}(e) exhibits a symmetric double hump structure with a global minimum associated with the peak in the time-integrated concentration. The hump in this EPR profile is related to the secondary peak in the concentration profile and acts as a marker for transition from incoherence to coherence regime. Hence, the transition between coherence and incoherence regime costs the most spatial increase in dissipation. This kind of regular structure corresponding to the incoherence state of the chimera is reminiscent of the previously demonstrated highly ordered time-average states of the chaotic spatiotemporal pattern~\citep{taodrder}.     
	\begin{figure*}
		\includegraphics[width=\textwidth]{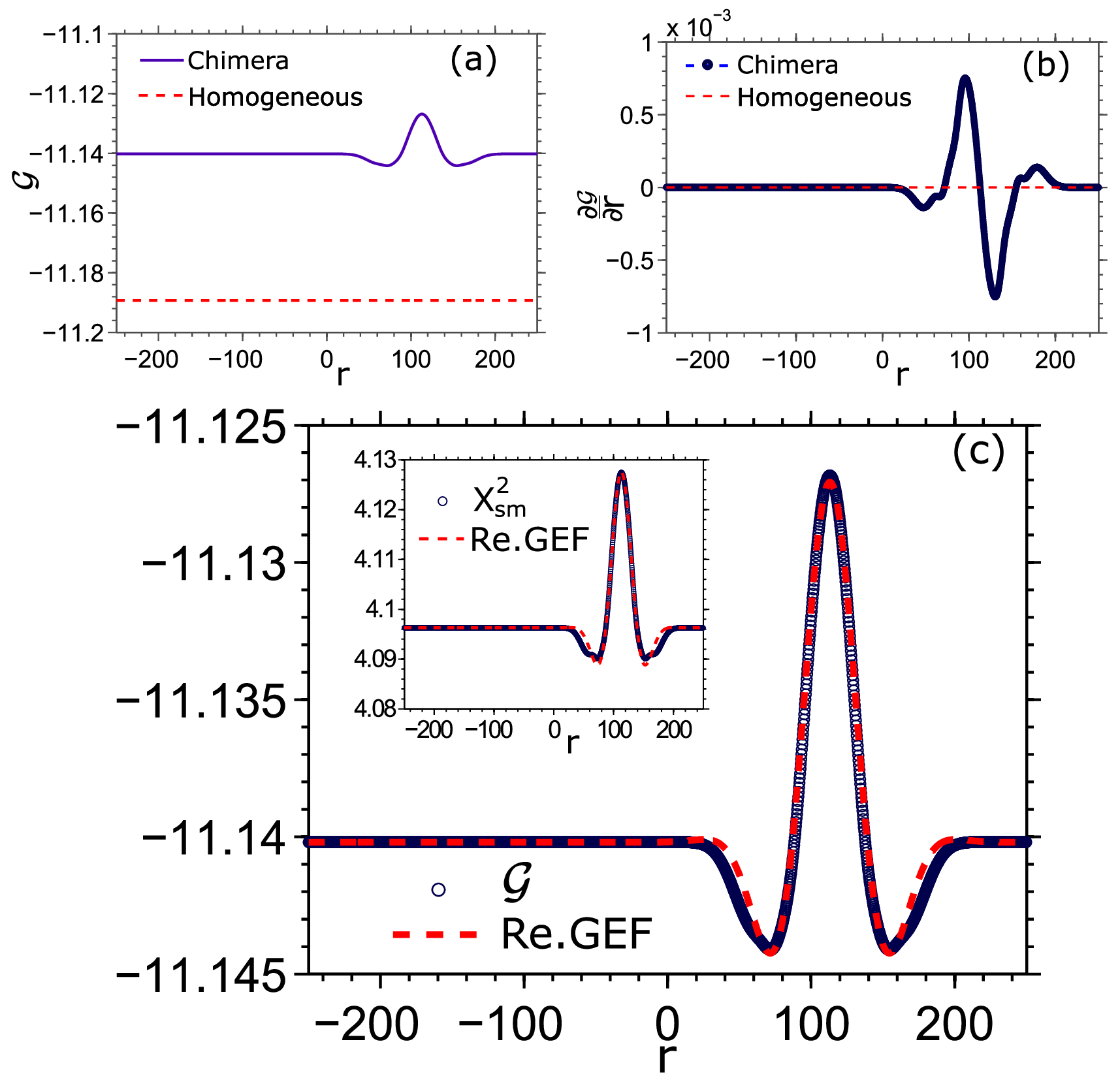}
		\caption{\label{energy} The semigrand Gibbs free energy(SGG) of chimera (a) The SGG of the chimera state (solid blue line), and corresponding (b) slope(blue line-circular dot) compared to the same entities for the homogeneous state(red `dotted' line). The homogeneous counterpart refers to a uniform field set by steady-state values of intermediate species. (c) The real part of the Gabor elementary function(GEF) is fit to the semigrand Gibbs free energy. Chemostatted species pair, $A$  and $B$ have been considered as the reference chemostatted species. All parameters are as in fig.~\ref{concentration}.}
	\end{figure*}
	
	Figure~\ref{energy}(a) demonstrates that the semigrand Gibbs free energy of the chimera state is always greater compared to its homogeneous counterpart as they are connected by a non-negative relative entropy of concentration distributions~\citep{Falasco2018InformationPatterns}.The homogeneous counterpart is related to a uniform concentration field of the system set by steady-state values of intermediate species. The nature of the transition among coherence and incoherence states in chimera is quite apparent in terms of energetics of the system illustrated through the semigrand Gibbs free energy profile and its spatial gradient in fig.~\ref{energy}(a), and ~\ref{energy}(b), respectively. More specifically, a notch on both sides marks the transition from coherence to incoherence state, and this notch is generated due to the previously mentioned maximum spatial dissipation rate during the transition. Additionally, the core of the incoherence state in this chimera is identified by a symmetric peak which hints at the energetically less stability of the incoherence state than its coherence counterpart.
	
	More interestingly, we have identified that the semigrand Gibbs free energy characteristics of the chimera render a 1D Gabor elementary function(GEF)~\citep{gabor1946theory} described as a Gaussian-modulated complex exponential function,
	$\exp{(-\frac{r^2}{2\tilde{w}^2}-i\tilde{k}r)}$, with the standard deviation,  $\tilde{w}$ of the Gaussian envelope, and preferred wavenumber, $\tilde{k}$. Thus, the semigrand Gibbs free energy here can be equivalently expressed in terms of a scaled and translated improved family of the Gabor elementary function as 
	$\mathcal{G}\equiv\mathcal{G}_c+Q\exp{(-\frac{(r-r_0)^2}{2\tilde{w}^2}-i\tilde{k}(r-r_0))}$, where $Q$ is the scaling factor, $r_0$ is the location of the center, and $\mathcal{G}_c$ is the semigrand Gibbs free energy of coherence regimes acting as the shift factor. The family of Gabor elementary functions in the above expression maintains the lowest possible bound of joint uncertainty in space and wavenumber, $\sigma_r\sigma_k=\frac{1}{2}$. 
	\begin{center}
		\begin{table}[h!]
			\begin{tabular}{|c c c c c c|} 
				\hline
				b & $\mathcal{G}_c$& $(Q, \tilde{w})$  & $(r_0,\tilde{k})$ & $(\sigma_r\sigma_k)_c$& $(\sigma_r\sigma_k)_s$\\  
				\hline
				5.24 & -11.11&(0.011, 37.50) & (137.70, 0.063) & 2.36 & 2.46\\
				\hline
				5.28 & -11.14& (0.013, 30.79)& (113.04, 0.060) &  1.68 & 2.14\\
				\hline
				5.32 & -11.17 & (0.017, 30.50) & (112.30, 0.0485) & 1.14 & 2.00\\
				\hline
				5.38 & - 11.21 &(0.024, 27.50) & (111.82, 0.043) & 0.82 & 1.90\\ 
				\hline
			\end{tabular}	
			\caption{ Joint uncertainty metric of the real, $(\sigma_r\sigma_k)_c$ and imaginary, $(\sigma_r\sigma_k)_s$ components of Gabor elementary function for different  values of control parameter, b.}
			\label{table:1}
		\end{table}
	\end{center}
	
	Due to the even-symmetry of the semigrand Gibbs free energy characteristics in fig.~\ref{energy}(a), the real component of the 1D Gabor elementary function with a suitable choice of parameters provides a reasonable fit to the semigrand Gibbs free energy profile as shown in fig.~\ref{energy}(c). The intriguing agreement of the chimera energetics with the Gabor elementary function allows one to exploit the interpretability of the Gabor elementary function to predict and manipulate the information transmitting capacity of the chimera over a given spatial domain. For example, one can readily say that Gabor's uncertainty principle~\citep{gabor1946theory, Farge, Daugman85} for information regarding localization trade-off in two conjugate domains is also equally valid for the semigrand Gibbs free energy of the chimera state. Moreover, the association of the semigrand Gibbs free energy with the Gabor elementary function also means that different chimera energy profiles at separate parametric regimes can be generated by dilation and translation of a particular Gabor elementary function related to a specific semigrand Gibbs free energy structure. Thus corresponding to energetic responses of chimera states at various control parameter values, we have identified different preferred wavenumber,$\tilde{k}$, center location, $r_0$, scaling factor,$Q$ and Gaussian envelope width, $\tilde{k}$ (see table \ref{table:1}) of Gabor elementary functions having spatial-wavenumber localization trade-offs. Interestingly, it is evident from the table \ref{table:1} that, unlike the whole Gabor elementary function case, the real and imaginary components of Gabor elementary functions have a Gaussian width-dependent joint uncertainty metric that can serve as an independent marker of chimera energetics.
	
	The Gabor representation of chimera energetics reveals the role of the constraint put by the uncertainty principle of information in shaping the thermodynamic evolution of the system. Besides these, the Gabor elementary function quantification of the time-integrated chimera dynamics provides crucial knowledge about evaluating the chimera state from an information-theoretic viewpoint; for example, the wavenumber selectivity for the coherent profile can be estimated from the center wavenumber($\tilde{k}$) of the Gabor elementary function fitted to the time sample-averaged $X^2$(inset in fig.~\ref{energy}(c)).     
	
	\section{\label{secVI}Conclusions}
	To sum up, our work has captured the emergence of chimera within a globally coupled continuum chemical oscillatory system and presented the corresponding nonequilibrium thermodynamic signatures. On this basis, we have identified the association of the chimera energetics with Gabor representation having the minimum theoretically possible joint uncertainty metric. Our demonstration of beat characteristics in the temporal rate of entropy production and the symmetric profiles of spatial entropy production in this study can be treated as key diagnostic elements to detect the nature of chimeras under different coupling schemes and diverse collective systems. In this regard, the thermodynamic characterization of chimera can be extended to traditional chimera in a ring of nonlocally coupled oscillators~\citep{Abramsring} or multi-chimera states resulting from strong nonlinear coupling~\citep{Omelchenko}. The thermodynamic insight of the chimera state can also be applicable in investigating the connection between the chimera state and synchronous state in Kuramoto-type networks~\citep{Chimerakuramotomodel}. For instance, these thermodynamic signatures can be utilized to qualitatively differentiate the chimera from other symmetry-breaking phenomena of such networks. Hence, detailed comparative studies aiming at the entropic and energetic signature for various chimera  classes~\citep{classichimera, PARASTESH20211} need to be carried out.
	Moreover, rendering the energy characteristics of the chimera in terms of Gabor representation here would complement the information thermodynamic~\citep{Falasco2018InformationPatterns} aspect of the pattern formation. The Gabor correspondence of the chimera energetics would immensely aid in accessing the similarity of the chimera with various other seemingly analogous states~\citep{chowbumps, RATTENBORG2000817, barkley, Duguet} and thus provide crucial insight into the understanding of the general spatial pattern of partial synchrony. The thermodynamic quantification of the chimera and its information-theoretic connection conferred here can enhance the efficiency of a chimera-based architecture~\citep{kanikabrain} and may open up application possibilities of the state outside the laboratory, from image representation ~\citep{1dgaborapplication, gaborapplication} to the representation of the primary visual system~\citep{DAUGMAN1980847}.   
	
	\section*{Acknowledgments} P.K. acknowledges fruitful discussions with Katharina Krischer and Sindre W. Haugland concerning numerical aspects of chimera states. 
	
	\bibliography{Chimera_letter_Ref}
\end{document}